# MUSIC, COMPLEXITY, INFORMATION
*Damián Horacio Zanette – April 2008*

"*Two impulses struggle with each other within man: the demand for repetition of pleasant stimuli, and the opposing desire for variety, for change, for a new stimulus.*" With these words, Arnold Schoenberg introduced the two fundamental principles which cast musical form. Repetition of perceptual elements – melodic motifs, rhythmic patterns, harmonic progressions– brings coherence to the musical structure, which is the basis of its comprehensibility. Variation, in turn, is necessary to avoid monotony and dullness. "*Variety is the mother of delight in Music,*" as Schoenberg's fellow composer Giovanni Maria Bononcini had put it three centuries earlier. In the search for a satisfactory combination of intelligibility and aesthetic substance, music conveys a subtle balance of reiteration and change, of redundancy and novelty, of recurrent shapes and fresh figures.

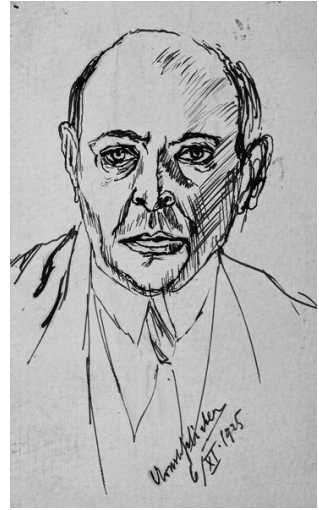

*Arnold Schoenberg (Selfportrait, 1925) conceptualized the complementary roles of repetition and variation in musical composition.*

This delicate equilibrium between uniformity and diversity evokes the nature of the class of entities that we call *complex systems*. An intermediate degree of internal organization, which makes a system elude incoherent behaviour but, at the same time, allows for rich dynamics and functional flexibility, is the key signature of complexity. The disordered, random-like motion of gas molecules, or the periodic, unrelenting ticking of a clock, hardly qualify as a complex system's outcome. On the other hand, even the most elementary function of a modest bacterium reveals the underlying complexity of the living organism.

Is it possible to quantify the degree of complexity of a system by measuring its distance to both randomness and order? A first empirical answer to this question was advanced in the 1930s by the philologist G. K. Zipf, who studied the statistics of word repetitions in transcriptions of long speeches and in written texts. He discovered a strong regularity in the relative frequencies of word occurrences –now known as *Zipf's law*– which is vastly widespread over different authors, styles, and languages: the number of repetitions of the $n$-th most used word is approximately proportional to the inverse of the rank $n$. If, for instance, the 10th most used word in a text occurs 300 times, Zipf's law predicts that the 100th most used word will appear some 30 times. From the work by M. G. Boroda, B. Manaris, and collaborators, we now know that a regularity similar to Zipf's law holds for musical compositions. In this case, the role of words in the statistics of repetitions can be replaced by single notes –each one defined by its pitch and duration– or composite items such as note duplets and triplets, interval successions, and chords.

Two decades after Zipf's work, social scientist H. Simon pointed out that Zipf's law can be quantitatively explained by assuming that, as a text is generated, the usage frequency of a word increases proportionally to its previous appearances. This very simple dynamical rule for the reinforcement in the usage of words during text generation, which statistical mathematicians call a *multiplicative process*, was enough for Simon to derive the inverse relation between number of occurrences and rank. More recently, it has been shown that the same rule explains the relative frequency in the usage of single notes in musical pieces, suggesting a strong affinity between the process of text generation and music composition.

Simon's model can be conceptually interpreted as representing the progression of the author's choices along the creative process –grammatical, morphological, semantic in language; melodic, harmonic, dynamical in music– which shape the work's intelligibility. It captures a basic mechanism of reinforced use of certain perceptual elements, whose recurrence is essential to elicit lasting neurophysical and psychological responses in the listener's brain, from creation and evocation of memories to association with images, sensations, and feelings. Of course, real literary texts and musical compositions are created as organic entities, not just as series of isolated decisions. The outcome, nevertheless, is an ordered sequence of events conveying information –a meaningful message. As the message flows, a cognitive frame supporting its coherence –the *context*– emerges spontaneously, favoring in turn the appearance of some elements at the expense of others. From this viewpoint, Simon's model for Zipf's law furnishes a unification of the concept of context in language and music. At the same time, variations in the form of Zipf's law for musical compositions of different authors make it possible to discern between their specific choices. The intentional lack of definition of a tonality context in Schoenberg's atonal pieces, for instance, can be quantitatively compared with the more consistent, but less flexible, use of tonal perceptual elements by Bach or Mozart. It is nevertheless suggestive that the dodecaphonic technique devised by Schoenberg to compose tonality-devoid music, is still based on the principles of repetition and variation, in this case, of prescribed twelve-tone sequences.

The statistical characterization of the structure of music is not exhausted by Zipf's law, though. Suppose that all the notes of a composition are shuffled and rearranged at random, preserving however the number of occurrences of each note. Zipf's law would still hold, but the musical message would go completely lost! This "thought experiment" calls attention on the fact that, besides the frequency of repetition of notes, their organization inside the musical piece is essential to its comprehensibility. Their specific sequence and the variation of their relative frequencies along the composition define, at scales of increasing duration, characteristic intervals, melodic motifs, phrases, modulation passages and tonality domains, up to entire sections.

Information theory provides us with a variety of tools to statistically analyze a symbolic sequence. Among them, *segmentation* aims at detecting portions of the sequence which mutually differ as much as possible in their relative contents of the various symbols. It proceeds by steps, first dividing the whole sequence into two segments with a maximal difference in the frequencies of different symbols, and then iterating the algorithm on the resulting segments. The product is a dissection of the sequence into domains which are maximally divergent with respect to their composition. Segmentation was recently used to analyze the compositional structure of a musical piece, the first movement of Mozart's keyboard sonata in C major (K. 545), as a sequence formed by the twelve tones of the chromatic scale. Remarkably, all the segmentation steps detected significant variations in the tonal context of the piece. The first step separated the first section of the sonata –the exposition– from the remaining of the piece, at the precise point where the tonality abruptly changes from G major to G minor. The second step separated the first theme, in C major, from the second theme, in G major, at the middle of the first section, and detected the beginning of the final episode, where the second theme is re-exposed in C major. Further steps defined new segmentation points associated with the use of increasingly elaborated tonal resources, such as the modulation episode from the first to the second subject, a rather long sequence of fifths and, towards the end of the piece, the use of a diminished seventh cord. In short, segmentation revealed essentially the same tonal domains that a human, trained in musical analysis, would have pointed out. As demonstrated experimentally by C. Krumhansl on Mozart's sonata K. 282, segmentation of musical compositions done by non-specialist listeners also occurs at the "tension peaks" determined, among other factors, by tonality changes.

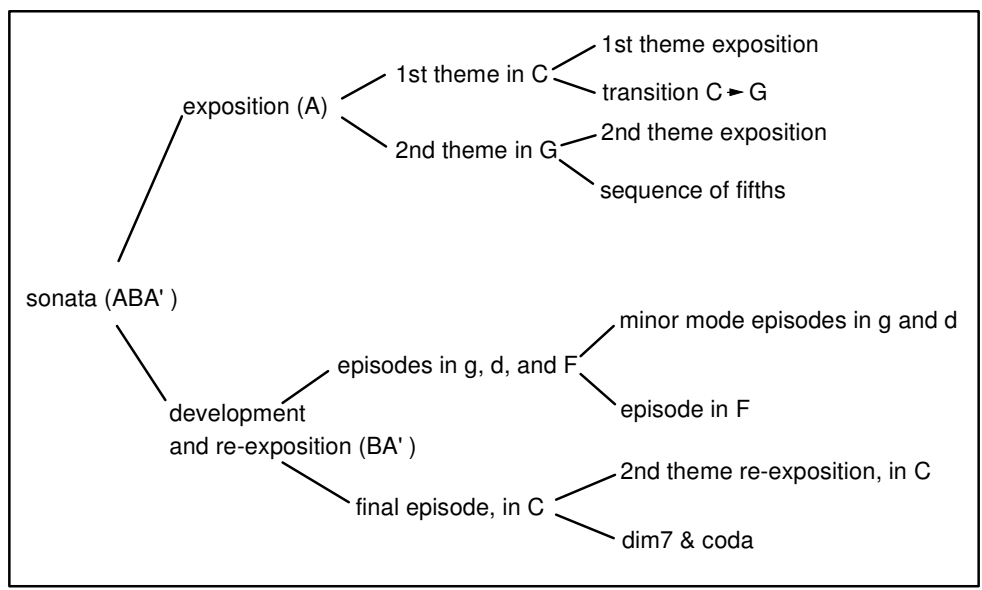

*Result of the three first segmentation steps on the first movement of Mozart's sonata K. 545, starting from the traditional ternary (sonata, ABA') form of this composition.*

The segmentation algorithm is not necessarily restricted to the analysis of the relative distribution of the twelve tones of the musical scale. A musical message can be interpreted as a sequence formed by other collections of symbols, for instance, combinations of pitch and duration, notes endowed with dynamical attributes, intervals, and chords. Using these more complex items, segmentation may disclose patterns related to richer cognitive qualities of music, such as those which the listener associates with mood and affect. However, as the collection of symbols grows in size and sophistication, their individual frequencies in the musical message decrease. Eventually, the algorithm will encounter a limitation which pervades all statistical methods –finite sampling. When the number of occurrences of each symbol ceases to be statistically significant, a meaningful message cannot be distinguished from a random sequence.

In the context of DNA sequence analysis, where segmentation has been used to detect long-range correlated patterns, it has been suggested that a broad distribution of segment lengths –with a substantial number of long, intermediate, and short segments– may function as a quantifiable signature of complexity. Indeed, upon segmentation, both random and periodic symbolic sequences yield little variation in the segment lengths. The complex information-carrying DNA sequence, on the other hand, is characterized by a long-tailed, slowly decaying distribution. Testing this suggestion on other symbolic sequences, such as those of linguistic or musical origin, is at this moment a fully open issue. It is difficult to overestimate the interest of procuring an efficient *complexity measure*. Such a tool –which should enter the category of universal quantifiers of the type of information and entropy– would be used to objectively settle some up to now sterile discussions regarding the relative complexity of different human languages or musical styles.

Besides contributing to elucidate the organizational structure of music, and its connection with the cognitive processes it induces, statistical methods of musical analysis may find more mundane applications in tasks related to music classification, such as authorship attribution, and style or period recognition. Automated classification of literary texts –a well developed branch of applied computer sciences– is based on the simultaneous use of several algorithms, which analyze from grammar constructions and syntax, to long-scale correlations and relative frequencies in word usage. The explosive growth of electronic text databases requires high-performance classification algorithms, able to avoid allocation errors which would amount to misplace a book in a huge library. In contrast, the problem of automatically classifying music is, at most, at an incipient stage. Somehow paradoxically, while –due to their lack of functional semantics– the structure of musical messages is much more flexible and diversified than that of literary texts, the identification of the key traits that define such diversity is much more elusive. For instance, hardly anyone who is minimally acquainted with classical music will mix up a concert by Vivaldi with one by Haydn, even when they were composed just three decades apart. But precisely what are the most significant features which make them so obviously different?

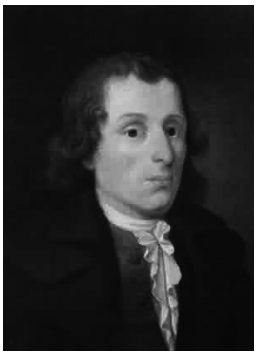

*Andrea Luchesi (1741-1801). The real composer of many of Mozart´s symphonies?*

Music-classification algorithms based on statistical methods may not only help to automate the organization of electronic databases, but could also advance a solution to long-dating questions of authorship attribution, giving at the same time confidence bounds for their answers: Is the score of *L'Incoronazione di Poppea* that we know today the product of Monteverdi's inspiration, or the collaborative work of several editors during its early performances all over Italy? Did Bach really write the choral *Nun ist das Heil und die Kraft* (BWV 50), whose autograph was never found? Was Andrea Luchesi the composer of many of the symphonies which, by a German prince's designs, were later attributed to Mozart?

The impersonal, emotionless, objective nature of statistical analysis seems to be at odds with many traditional ways of pondering a work of art, which emphasize almost ineffable aesthetic nuances, psychological qualities, and individual, intimate values. In fact, the integration and elaboration of sensory information into what we call an artistic experience, in all its intricacy, may lie forever beyond the scopes of a mere quantitative description. On the other hand, quantitative methods can bring about and illuminate new facets of artistic creation, related to the organization of its many intermingled patterns into a comprehensible structure, thus enriching and complementing our still exploratory understanding of such a complex phenomenon.

Damián H. Zanette conducts research at the Statistical and Interdisciplinary Physics Group of Centro Atómico Bariloche, and is an associate professor of physics at Instituto Balseiro, Bariloche, Argentina.